\documentclass[twocolumn]{aa}
\usepackage{amssymb}
\usepackage{latexsym}

\usepackage{graphicx}

\title{Big-bang nucleosynthesis in Brans-Dicke cosmology \\
with a varying $\Lambda$ term related to WMAP
}
\author{R. Nakamura\inst{1} \and
M. Hashimoto\inst{1} \and S. Gamow\inst{1} \and K. Arai\inst{2}}

\institute{
Department of Physics, Kyushu University, Fukuoka, 810-8560, Japan \and
Department of Physics, Kumamoto University, Kumamoto, 860-8555 Japan 
}

\date{Received / Accepted}

\abstract{ 
We investigate the big-bang
 nucleosynthesis in  a Brans-Dicke model 
 with a varying $\Lambda$ term using the Monte-Carlo method and
 likelihood analysis. It is found that the
cosmic expansion rate differs appreciably 
from that of the standard model. 
 The produced abundances of $^4$He, D, and barely Li 
 are consistent with the observed ones
 within the uncertainties in nuclear reaction rates
 when the baryon to photon ratio 
 $\eta = \left( 5.47 - 6.64 \right)\times 10^{-10}_{}$, which is in 
 agreement with the value deduced from WMAP.
\keywords{general:nuclear reactions -- nucleosynthesis -- abundances --
cosmology:early universe -- cosmic microwave background}
}

\begin{document}

 \titlerunning{ Big-bang nucleosynthesis in Brans-Dicke cosmology with a
 varying $\Lambda$ term}
 \authorrunning{Nakamura  et al.}

\maketitle

%

\section{Introduction}

The standard model of the
big-bang nucleosynthesis (SBBN) has
succeeded in explaining the origin
of light elements $^4_{}$He, D, and $^7_{}$Li. 
Although the value of the baryon-to-photon ratio $\eta$
has been derived from 
the observations of the Wilkinson Microwave
 Anisotropy Probe (WMAP) (Bennett et al. \cite{Bennett})
to be ~$\eta^{}_{10}=6.1^{+0.3}_{-0.2}$
, the value seems to be inconsistent
 with the results of SBBN (Coc et al. \cite{coc}). 
 Contrary to the excellent concordance with $\eta$ of
 WMAP for D, the abundance of $^4_{}$He by SBBN is rather low compared
 to that from WMAP. Therefore, non-standard models of BBN have been
 proposed with the Friedmann model modified (Steigman \cite{Steigman}).

 For non-standard models, scalar-tensor theories have been investigated
 (e.g., Bergmann \cite{Bergmann}; Wagoner \cite{Wagoner}; 
Endo \& Fukui \cite{Endo};
Fukui et al. \cite{Fukui}). For a simple model with a scalar $\phi$, it is
 shown that
 a Brans-Dicke (BD) generalization of gravity with torsion includes the low-energy
limit string effective field theory (Hammond \cite{Hammond}).
Related to the {\it cosmological constant problem},
 a Brans-Dicke model with a varying $\Lambda(\phi)$ term (BD$\Lambda$)
has been presented, 
and also investigated from the point of an inflation theory (Berman \cite{Berman}). 
Moreover, it is found that the linearized 
gravity can
 be recovered in the Randall-Sundrum brane world 
(Garriga \& Tanaka \cite{Garriga}).
Furthermore, scalar-tensor cosmology is constrained by $\chi^2_{}$ test
 for WMAP spectrum (Nagata et al. \cite{nagata}) where
the present value of the coupling parameter $\omega_0 =\omega(\phi_0)$
is bounded to be  $\omega^{}_0 > 50~(4\sigma)$
 and $\omega^{}_0 > 1000~(2\sigma)$ in the limit to BD cosmology.

In the mean time, BBN has been studied 
in BD$\Lambda$ (Arai et al. \cite{Arai}; Etoh et al. \cite{Etoh}). 
A relation between BBN and scalar-tensor gravity is investigated with the
inclusion of $e^{+}e^{-}$ annihilation to the equation of state, where
the present value of the scalar coupling has been constrained
(Damour \& Pichon \cite{damour}).
On the other hand, it is suggested that a decaying $\Lambda$ 
modifies the evolution of the scale factor and
affects the temperature  $T_r$ of the cosmic microwave background 
at redshift $z \le 10^4$, when the 
recombination begins due to the decrease in $T_r$
(Kimura et al. \cite{Kimura}), 
while
a decaying $\Lambda$ is found to be consistent with temperature observations 
of the cosmic microwave background
for $z < 4$ (Puy \cite{Puy}). 
Therefore, it is worthwhile to check the validity of
BD$\Lambda$ related to the recent observations.
In the present paper, we investigate how extent 
 BBN in the BD$\Lambda$
model can be reconciled with $\eta$ from WMAP.

In \S 2, the formulation for BD$\Lambda$ is given
and the evolution of the universe in BD$\Lambda$ is shown.
Our results of BBN are presented in \S 3 using
the Monte-Carlo method~(Cyburt et al. \cite{Cyburt}),
and 
constraints are given to the parameters inherent in
BD$\Lambda$.
We examine in \S 4 the evolution of the scale factor 
and the resulting abundances
with taking into account the deviation from the equation of state
$p = \rho/3$ during the stage of $e^{+}e^{-}$ annihilation.
In \S 5, the likelihood analysis~(Fields et AB. \cite{Fields}) 
is adopted to get most probable values and the accompanying errors.

\section{Brans-Dicke cosmology with a varying $\Lambda$ term}
\begin{figure*}
\centering
 \includegraphics[width=12cm]
 {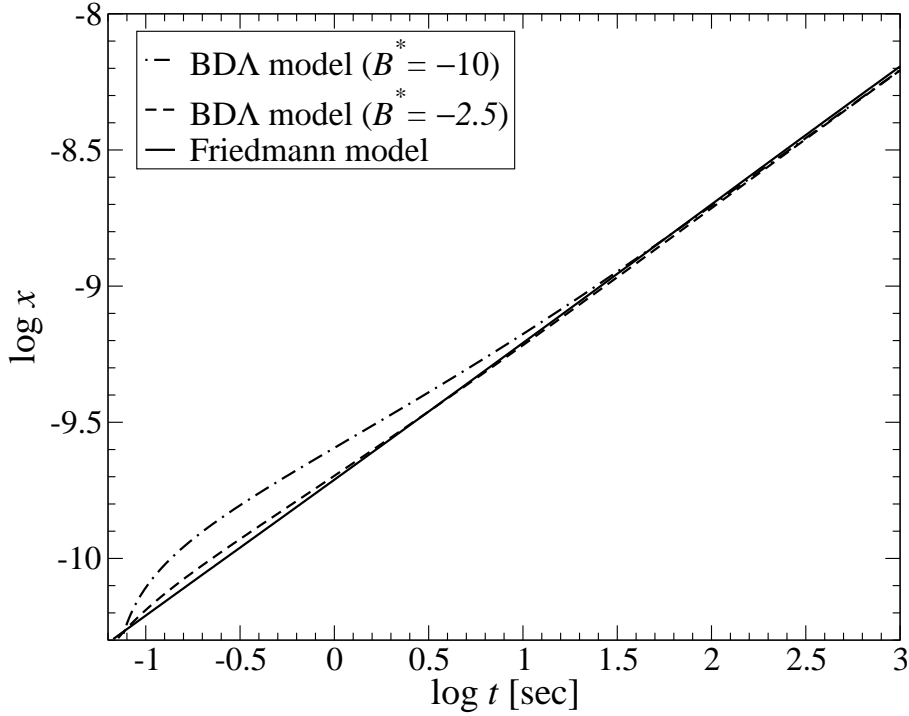}
 \caption{Evolution of the scale factor 
 for BD$\Lambda$ ($\mu=0.7$, $t^{}_0 =13.7$~Gyr, and $\eta^{}_{10}=6.1$)
 and Friedmann model.} \label{figx}
\end{figure*}

The field equations for BD$\Lambda$ are written as follows 
(Arai et al. \cite{Arai}):
   \begin{eqnarray}
    R_{\mu\nu} -\frac{1}{2} g_{\mu\nu} R + g_{\mu\nu} \Lambda
     \hspace{-1mm}&=&\hspace{-1mm} \frac{8\pi}{\phi} T_{\mu\nu} + \frac{\omega}{\phi^2}
     (\phi_{,\mu;\nu} - \frac{1}{2} g_{\mu\nu}\phi_{,\alpha}\phi^{,\alpha}) \nonumber \\
     &+&\hspace{-1mm} \frac{1}{\phi}(\phi_{,\mu;\nu} - g_{\mu\nu} \Box\phi)
     \label{field1},
   \end{eqnarray}
   \begin{equation}
      R-2\Lambda-2\phi\frac{\partial\Lambda}{\partial\phi}=
      \frac{\omega}{\phi^2}\phi_{,\mu}\phi^{,\mu}-\frac{2\omega}{\phi}\Box\phi
      \label{field2},
   \end{equation}
where $\omega$ is the coupling constant. 

The equation of motion is obtained with use of the
Friedmann-Robertson-Walker metric:
\noindent
\begin{equation}
 ds^2 = -dt^2 + a(t)^2\left\{ \frac{dr^2_{}}{1-kr^2_{}} +
		       r^2_{}d\theta^2_{}+ r^2_{} \sin^2\theta d\phi^2
		\right \},
\end{equation}
where $a(t)$ is the scale factor and $k$ is the curvature constant.
Let  $x$ be a scale factor normalized to its  present value, 
i.e., $x=a/a_0$,
then we get from the $(0,0)$ component in Eq.~(\ref{field1})
   \begin{equation}
    \left(\frac{\dot x}{x} \right)^2_{} +
     \frac{k}{x^2_{}}-\frac{\Lambda}{3}
   -\frac{\omega}{6}\left( \frac{\dot{\phi}}{\phi} \right)^2_{}
   -\frac{\dot x}{x}\frac{\dot \phi}{\phi}
   = \frac{8\pi}{3}\frac{\rho}{\phi} \label{eqom} , 
   \end{equation}
where $\rho$ is the energy density.

We assume the simplest case of the coupling between the scalar and 
matter fields:
   \begin{equation}
      \Box\phi =
         \frac{8\pi}{2\omega + 3}\mu T_{\nu}^{\nu} \label{sm1} \,,
   \end{equation}
   where $\mu$ is a constant.
   Assuming the perfect fluid for $T^{}_{\mu\nu}$, Eq.~(\ref{sm1}) reduces to
   \begin{equation}
    \frac{d}{dt}(\dot\phi x^3)=\frac{8\pi\mu}{2\omega+3}(\rho-3p) x^3_{}
     \label{sm2} ,
   \end{equation}
   where $p$ is the pressure. 

   A particular solution of Eq.~(\ref{field2}) is obtained from
   Eqs.~(\ref{field1}) and (\ref{sm1}):
   \begin{equation}
    \Lambda =\frac{2\pi\left( \mu-1
		       \right)}{\phi}\rho_{m0}x^{-3}
    \label{lambda},
   \end{equation}
where $\rho_{m0}$ is the matter density at the present epoch.

The gravitational ``constant'' $G$ is expressed as follows
   \begin{equation}
    G=\frac{1}{2}
     \left(3-\frac{2\omega+1}{2\omega+3} \mu\right)
     \frac{1}{\phi}
     \label{graphi}  .
   \end{equation}

The radiation density $\rho_r$ contains the contributions from photons, neutrinos,
electrons and positrons at $t\le 1$~s. The total energy density is given as
\begin{eqnarray}
 \rho & = & \rho^{}_m + \rho^{}_r, \ \  \rho^{}_r =\rho_{rad}+\rho_{\nu}+\rho_{e^{\pm}}
  \label{rhop}.
\end{eqnarray}
Here the energy density of matter varies as $\rho^{}_m=\rho^{}_{m0}x^{-3}_{}$.
The radiation density $\rho^{}_r=\rho^{}_{r0}x^{-4}_{}$ except $e^{+}e^{-}$ epoch
where $e^{+}e^{-}$ annihilation changes the relation $T_r \sim x^{-1}${}.
We assume that the pressure satisfies $p=\rho/3$, which is legitimated only for relativistic
particles. Then, Eq.~(\ref{sm2}) is integrated to give
\begin{equation}
 \dot\phi = \left(\frac{8\pi\mu}{2\omega + 3}\rho^{}_{m0}t+B\right)
  \frac{1}{x^3_{}} \label{phit},
\end{equation}
where $B$ is a constant (Arai et al. \cite{Arai}).
Although the relation $p = \rho/3$ does not
hold during the epoch of $e^{+}e^{-}$ annihilation,
as pointed out by Damour and Pichon (1999),
the inclusion of $\phi$ measures small
deviation from SBBN in the case of our interest,
so that our solution (\ref{phit}) can reasonably describe 
the evolution of $\phi$ except the annihilation epoch. We consider that $B$
affects the evolution of $x$ significantly from the early epoch to the present
compared to the contribution from $e^{+}e^-$ annihilation. As the consequence,
neutron to proton ratio is affected seriously by the initial value of $B$.
Considering the important contribution of $\rho_{e^{\pm}}$ to BBN, we examine
the effects of $e^{+}e^{-}$ annihilation in \S 4.
Hereafter we use the normalized values:
$B^* = B/(10^{-24}$ g s cm$^{-3}$), 
and $\eta_{10}=10^{10}\eta$.
The coupled equations ~(\ref{eqom}), ~(\ref{lambda}), and ~(\ref{phit})
can be solved numerically with the specified 
quantities: as for macroscopic quantities,
$G_0 = 6.672 \times 10^{-8}$ dyn cm$^2_{}$ g$^{-2}_{}$, 
$H_0=71$ km s$^{-1}$ Mpc$^{-1}$ (Bennett et al. \cite{Bennett}),
and $T^{}_{r0}=2.725$ K (Mather et al. \cite{Mather});
as for microscopic quantities, the number of massless neutrino
species is 3 and the half life of neutrons
is $885.7$ s (Hagiwara et al. \cite{Hagiwara}). Though we adopt $\omega=500$, 
the epoch of the appreciable growth of $|\dot{G}/G|$ is $t < 10^3$ s
regardless of the value $\omega$ (Arai et al. \cite{Arai}). Therefore,
even if we adopt a value $\omega> 500$ (Will \cite{will}),
we can get qualitatively the same conclusion by changing the parameters
$\mu$ and $B^*$. We impose the condition
 $|\dot{\phi}/\phi|^{}_{0}=|\dot{G}/G|^{}_{0}<10^{-13}_{}$ yr$^{-1}_{}$
 which is the most severe observational limit 
(M\"{u}ller et al. \cite{Muller}).
 
 We remark that BD$\Lambda$ is an extension of the original form of BD and reduces
 to the Friedmann model when $\phi$ = constant, $\mu = 1$, and $\omega \gg 1$.
 We have $\Lambda < 0$
 if $\mu < 1$, and $\phi G > 0$ if both $\mu > 3$ and $\omega \gg 1$. 
 Figure~\ref{figx} shows the evolution of the scale factor
 for BD$\Lambda$ with the relevant parameters in the present study
 and for the Friedmann model.
 Note that the difference in the expansion rate 
 at $t < 10$~s in BD$\Lambda$.
 In particular, around $t=5$ s, the curve $x$ in BD$\Lambda$ crosses that
 of the Friedmann model, which will have sensitive 
 effects on BBN.
 Since $\Lambda$ is proportional to $\rho_{m0}$, $\mu$ 
 affects the evolution of the
 scale factor around the present epoch.
 In our BD$\Lambda$ model, if $|B^*_{}|$ increases, the expansion rate 
 increases at $t<10-100$~s.
 It is remarked that the change in $G$ between the recombination and 
 the present epoch is less than
 0.05 ($2\sigma$) from WMAP (Nagata et al. \cite{nagata}), which is 
 consistent with
 BD$\Lambda$ since $|(G-G_0)/G_0|<0.005$ at $t>1$~yr.

 \section{Big-bang nucleosynthesis}
    \begin{figure}
       \includegraphics[width=\linewidth]{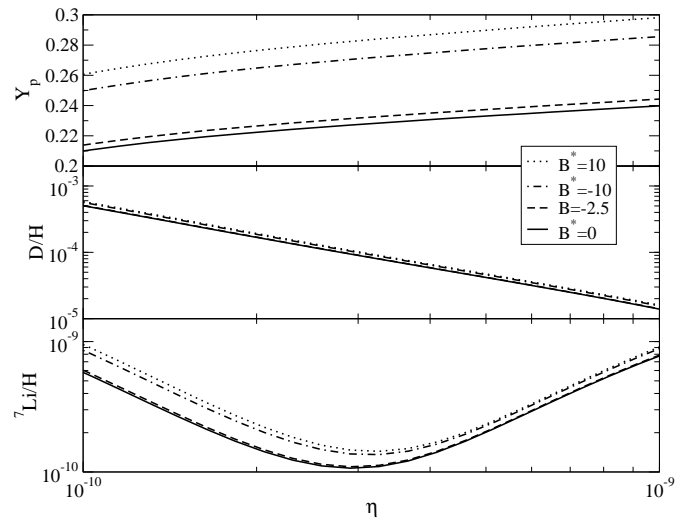}
       \caption{ Light element abundances against $\eta$ in BD$\Lambda$ for
       $\mu=0.7$ and possible values of $B^*_{}$. } 
     \label{fig:abn-bdl-b}
    \end{figure}
    \begin{figure}
     \includegraphics[width=\linewidth]{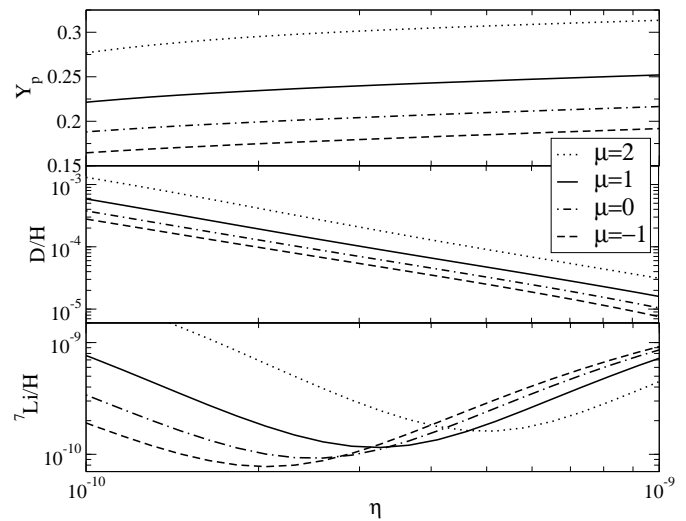}
     \caption{Same as Fig. \ref{fig:abn-bdl-b} but for $B^*_{}=0$ and
     various values of $\mu$. } \label{fig:abn-bdl-mu}
    \end{figure}
 
 Changes in the expansion rate compared to the standard model 
 affect the synthesis of light elements at the early era,  
 because the neutron to proton ratio is sensitive to the expansion rate.
 
 For the BBN calculation, we use the reaction rates~(Cyburt et
 al. \cite{Cyburt}) based on NACRE (Angulo et al. \cite{Angulo}). 
 We adopt the observed abundances of $^4_{}$He,~D/H and $^7_{}$Li/H
 as follows; 
 $Y_{\rm p}=0.2391\pm 0.0020$~(Luridiana et al.~\cite{Luridiana}), 
 D/H$=2.78^{+0.44}_{-0.38}\times 10^{-5}$~(Kirkman et
 al.~\cite{Kirkman}),
 ${}^7_{}$Li/H=$\left( 2.19\pm 0.28 \right) \times 10^{-10}_{}$
~(Bonifacio et al.~\cite{Bonifacio}). 

Since the results of WMAP constrain cosmological parameters,
we calculate the abundance of $^4$He, D and $^7$Li paying attention to
the value $\eta^{}_{10}=6.1$.
First, we carry out the BBN calculations with use of the adopted
experimental values of nuclear reaction rates given in NACRE.
Figure~\ref{fig:abn-bdl-b} illustrates $^4_{}$He, D/H, and $^7_{}$Li/H for
$\mu=0.7$.  The abundance of $^4_{}$He is very sensitive to both $B^{*}_{}$
and $\mu$; it increases if $|B^{*}|$ or $\mu$ increases.
On the other hand, D and $^7_{}$Li are more sensitive to $\mu$
than $B^{*}$ as seen from Fig.~\ref{fig:abn-bdl-mu}.
As a result, ${}^4_{}$He and D/H are consistent with $\eta$
obtained from WMAP in the range $-0.5\le\mu\le 0.8$ and 
$-10\le B^*\le 10$. 

\begin{figure}
  \includegraphics[width=\linewidth]{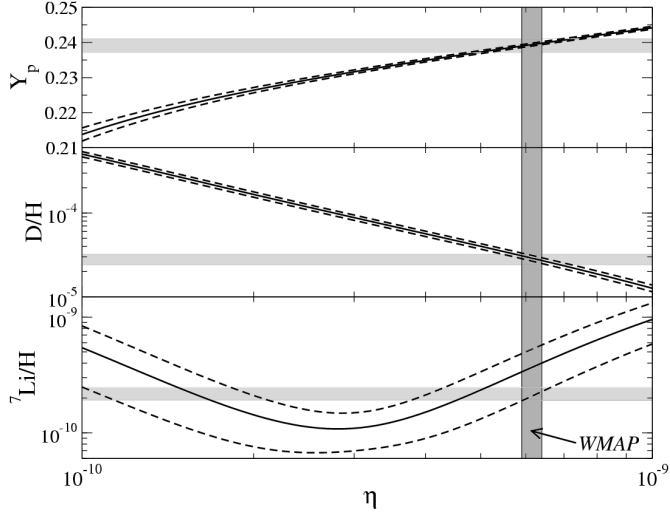}
  \caption{Light-element abundances vs. $\eta$ in BD$\Lambda$
  for $\mu=0.7$ and $B^*=-2.5$. Dashed lines
  show $\pm 2\sigma$ uncertainties in
  nuclear reaction rates. The dark-shaded area indicates the constraint by
  WMAP and light-shaded areas denote regions of observational abundances. }  
  \label{fig:eta-bdl}
 \end{figure}

Next, we perform 
the Monte-Carlo calculations to obtain the upper and
lower limits to individual
abundance using the uncertainties in the nuclear reaction rates
(Cyburt et al. \cite{Cyburt}). 
Figure \ref{fig:eta-bdl} illustrates 
$^4_{}$He,~D/H and ${}^7_{}$Li/H with $2\sigma$ uncertainties 
for $B^{*}_{}=-2.5$ and $\mu=0.7$.
The light-shaded areas denote the regions of observed 
abundances, and the dark-shaded area indicates the limit obtained from WMAP.
While the obtained values of $^4_{}$He and D are consistent with $\eta$
by WMAP, the lower limit in $^7_{}$Li is barely consistent.

%


\section{Effects of $e^{+}e^{-}$ annihilation on BBN}

In the previous sections, we have assumed the equation of state 
$p = \rho/3$ in Eq.~(\ref{sm2}) to obtain Eq.~(\ref{phit})
at the epoch of $e^{+}e^{-}$ annihilation. 
Let us discuss the effects of $e^{+}e^{-}$ annihilation on
the evolution of the scalar field and the scale factor
due to the deviation from the relation $p = \rho/3$.
The electron-positron pressure and energy density are written with the
variable $\zeta=m^{}_{e}/k^{}_{B}T^{}_r$ as follows

\begin{equation}
p^{}_{e} = \frac{2m_{e}^{4}}{ \pi^{2}_{} \hbar^{3}_{}} \sum_{n=1}^{\infty} (-1)^{n+1}_{}
\left( \frac{1}{n\zeta}\right)^{2} K_{2}\left( n\zeta \right),  \label{pee}
\end{equation}

\begin{equation}
\rho^{}_{e} = 3~p^{}_{e} + \frac{2m_{e}^{4}}{ \pi^{2}_{} \hbar^{3}_{}} \sum_{n=1} ^{\infty
} (-1)^{n+1} \left( \frac{1}{n\zeta}\right) K_{1}\left(
	      n\zeta \right) \, , \label{rhoee}
\end{equation}
where $\hbar$ is Planck's constant in units of $2\pi$, $k^{}_{B}$ is Boltzmann's constant, and
$m^{}_{e}$ is the electron rest mass.
$K_{i}\,(i=1 \mbox{ and } 2)$ are modified Bessel functions of order $i$
(e.g. Damour \& Pichon \cite{damour}). In the numerical calculations,
the summations in Eqs.~(\ref{pee}) and (\ref{rhoee}) are taken over $n=1-10$.
We can get the scale factor by
integrating Eq.~(\ref{eqom}) with the aid of Eq.~(\ref{sm2}).
To see the effects of $e^{+}e^{-}$, we take the form:
\begin{equation}
\dot\phi x^3= \frac{8\pi\mu}{2\omega + 3}\int^t{(\rho_e-3p_e)}x^3dt +B
\label{phix}.
\end{equation}

\begin{figure}
\includegraphics[width=\linewidth]{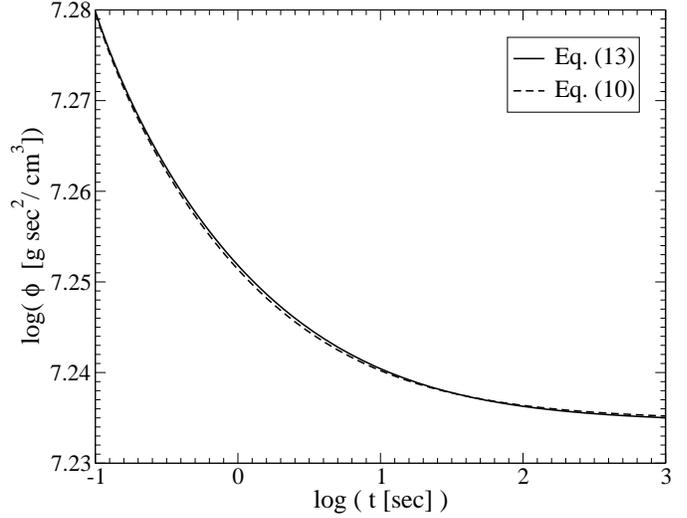}
   \caption{Evolution of the scalar field. The solid line refers 
the integration of Eq.(13) with $B^* = -2.43$,
and the broken line is for Eq. (10) with $B^* = -2.50$.} 
 \label{fig:phi_compare}
\end{figure}

A direct comparison is made for the evolution of the scalar field.
The results are shown in Fig.~\ref{fig:phi_compare}, where the solid line
indicates the case of Eq.~(\ref{phix}) with $B^{*}=-2.43$ and $\mu=0.7$,
and the broken line is the case of Eq.~(\ref{phit}) with 
$B^{*}_{}=-2.50$ and $\mu=0.7$.
These sets of parameters yield the same macroscopic quantities given in \S 2.
Although we can appreciate the slight difference at $t < 10^3$~s, 
it remains small during and after the stage of BBN.
The effects on the evolution of the scale factor are minor and
the change in $Y_p$ is found to be at most 0.1\% compared to that obtained in \S 3.
We can conclude that since the effects of $B$ in the range $-10\le B^*\le 10$ 
is much larger compared to those of 
$e^{+}e^{-}$ annihilation, 
the deviation from the relation $p = \rho/3$ due to $e^{+}e^{-}$ does not change our
results qualitatively. 
However, we note that even small differences in $Y_p$ may affect the detailed
statistical analysis combined with theoretical and observational
uncertainties performed in the previous sections.

\section{Discussion and conclusions}
 \begin{figure}
  \includegraphics[width=\linewidth]{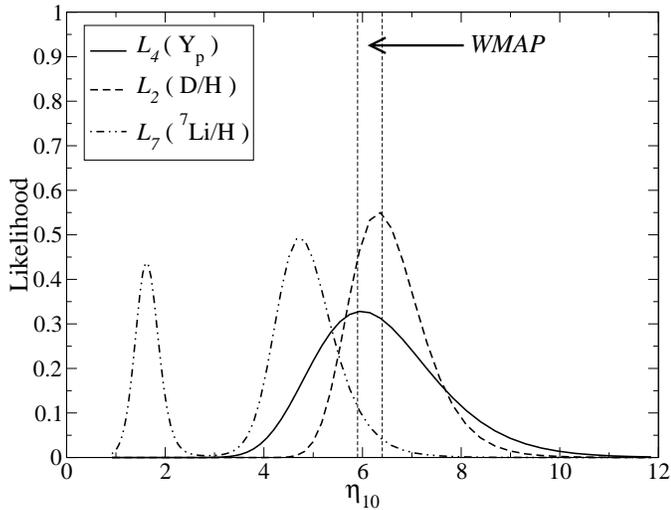}
   \caption{Likelihood function as a function of $\eta_{10}$ for
   $^4_{}$He~($L^{}_4$),~D~($L^{}_2$) and ${}^7_{}$Li~($L^{}_7$).
   The vertical lines indicate upper and lower limit to $\eta$ by WMAP.
   } \label{fig:lik-each}
\end{figure}

\begin{figure}
   \includegraphics[width=\linewidth]{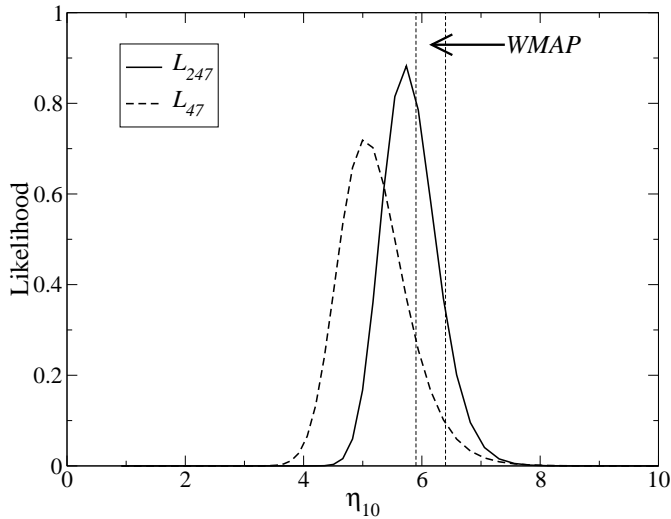}
   \caption{ Combined Likelihood function for two $(L^{}_{47})$ and
   three-elements $(L^{}_{247})$. } 
   \label{fig:lik-all}
 \end{figure}

We have carried out the BBN calculations in the $\mu-B^{*}$
plane and obtain 
the ranges $-0.5\le\mu\le0.8$ and $-10\le B^*\le10$ 
that are consistent with both the abundance observations and 
$\eta$ obtained from WMAP.

To evaluate uncertainties of theory and observations, we
calculate normalized
likelihood distributions in BBN (Fields et al.\cite{Fields}; Hashimoto
et al. \cite{Hashimoto}). 
In Fig.~\ref{fig:lik-each},~we show the likelihood functions for
${}^4_{}$He,~D, and ${}^7_{}$Li. The combined distributions,
$L^{}_{47}=L^{}_{4}\cdot L^{}_{7}$ and 
$L^{}_{247}=L^{}_{2}\cdot L^{}_{4}\cdot L^{}_{7}$ are shown in
Fig.~\ref{fig:lik-all}.
As the result, we get the 95\% confidence limit of $\eta$:
$ 5.47 \le \eta^{}_{10} \le 6.64 $.

The consistency holds within $1\sigma$ error for ${}^4_{}$He and D, and
$2\sigma$ for ${}^4_{}$He, D, and ${}^7_{}$Li. 
Though new reaction rates
recently published (Descouvemont et al. \cite{Descouvemont}) 
will change the errors to some extent
in the likelihood analysis, our conclusion holds qualitatively.

Our previous studies (Etoh et al. \cite{Etoh}) showed $1<\mu<3$ if $\Lambda>0$
for large value of $\omega$.
In the present case, the $\Lambda$ term becomes negative from
Eq.~(\ref{lambda}) for $\mu<1$: this would not conflict
with available observations and/or basic theory (Vilenkin \cite{vilen}).
Alternatively, if we consider $\Lambda=\Lambda_0 + \Lambda(\phi)$ with
$|\Lambda(\phi_0)/\Lambda_0| < 0.01$, then the cosmological term becomes
consistent with the present observations. Though the evolutionary path in the
early universe can deviate from the Friedmann model (Arai et al. \cite{Arai}),
parameters in BD$\Lambda$ must be searched in detail for values of
$\omega>500$ to get quantitative results of BBN.
We note that it is shown that negative energies are present in scalar-tensor
theories, though it is not clear how to identify them definitely (Faraoni \cite{fara}).

To save the apparent inconsistency for SBBN, effects of neutrino
degeneracy, changes in neutrino species, or other new physical processes
have been included (Steigman \cite{Steigman}). In our model, we need
only a scalar field that could be related to a string theory (Hammond \cite{Hammond}).
It is noted that the original BD 
 cosmology ($\mu=1$) would be limited severely by the more accurate
 observation of light elements and/or the future constraints for $\eta$
 as shown in the present investigation.

\begin{acknowledgement}
 Data analysis were in part carried out on a general common user computer system
 at the Astronomical Data Analysis Center of the National
 Astronomical Observatory of Japan.

\end{acknowledgement}

\end{document}